\documentstyle[12pt,aps,preprint,tighten,epsf]{revtex}

\begin{document}
\draft
\title{\begin{flushright}{\small Preprint CNS-9807\\Jan. 
1998}\end{flushright}
Nonadiabatic Geometric Phase for the Cyclic Evolution of
a Time-Dependent Hamiltonian System
}

\author{
Jie Liu $\,^{1,2}$,
Bambi Hu $\,^{1,3}$
, and Baowen Li$\,^1$ 
 }
\address{$^1$Department of Physics and Center for Nonlinear Studies,
\\ Hong Kong Baptist University, Hong Kong, China\\
$^2$Institute of Applied Physics and
Computational Mathematics\\ P.O.Box.8009,  100088 Beijing, China\\
$^3$Department of Physics, University of Houston, Houston,TX 77204 USA
}
\maketitle
\date{}
\begin{abstract}

The geometric phases of the cyclic states of a generalized harmonic
oscillator with nonadiabatic time-periodic parameters are discussed in the
framework of squeezed state.  It is shown that the cyclic and quasicyclic
squeezed states correspond to the periodic and quasiperiodic solutions of
an effective Hamiltonian defined on an extended phase space, respectively.
The geometric phase of the cyclic squeezed state is found to be a
phase-space area swept out by a periodic orbit. Furthermore, a class of
cyclic states are expressed as a superposition of an infinte number of
squeezed states. Their geometric phases are found to be independent of
$\hbar$, and equal to $-(n+1/2)$ times the classical nonadiabatic Hannay
angle. 

\end{abstract}
\section{Introduction}

Berry phase is one of the most important findings on the fundamental
problem of quantum mechanics in recent years\cite{Berry84}. It reveals the
gauge structure associated with a phase shift in adiabatic processes in
quantum mechanics. This quantal phase is connected with a classical angle,
namely, Hannay angle, by a simple and elegant expression in the
semiclassical limit\cite{HB85}. The next important progress has been the
relaxation of the adiabatic approximation\cite{AA87,BH88}. Aharanov and
Anandan studied the phase associated with a cyclic evolution in quantum
mechanics, which occurs when a state returns to its initial condition.
They 
shown that the phase is a geometrical property of the curve in the
projective Hilbert space which is naturally associated with the
motion$[3]$. For the special case of an adiabatic evolution, this phase
factor is a gauge-invariant generalization of the Berry phase. The
significance of Aharanov and Anandan's generalization lies in:  1) The
cyclic evolution of a physical system is of most interest in physics both
experimentally and theoretically; 2) The universal existence of the cyclic
evolution is guaranteed for any quantum system.  The second point can be
easily recognized by considering the eigenvectors of the unitary evolution
operator for a quantum system.  An explicit example is a time-periodic
Hamiltonian system where the Floquet theorem applies. The eigenfunctions
of the Floquet operator, which are so-called the Bloch wave functions in
the condensed matter physics, are obviously cyclic solutions and of great
interest in physics. However, unlike the adiabatic case, in the
nonadiabatic case, calculating the eigenvectors and extracting the
nonadiabatic geometric phase from the quasi-energy term for a
time-dependent Hamiltonan is far from an  trivial work, except for
such a special example as the spin particle in magnetic field.  Recent
works of Ge and Child\cite{GC97} made a step further in this direction.
They found a {\it special} cyclic state of Gaussian wave packet's form for
a generalized harmonic oscillator. The nonadiabatic geometric phase is
explicitly calculated and found to be one half of the classical
nonadiabatic Hannay angle. 

In this paper, we would like to suggest an alternative approach - the
squeezed state approach to study the {\it general} cyclic evolutions of
the generalized harmonic oscillator. In particular, we shall construct a
class of quantum states based on the superposition of an infinite number
of squeezed states. We find that the condition for them to be cyclic
evolutions is nothing but quantization rule without Maslov-Morse
correction. The nonadiabatic geometric phases are obtained {\it
analytically}, and found to be related to the classical Hannay angle by an
explicit expression. 
 
Squeezed state approach has wide applications in many branches of physics
such as quantum optics and high energy physics etc. Recent years has
witnessed a growing application of squeezed state to the study of chaotic
dymanical systems\cite{PS94,ZF95,LS97,HLLZ98}. In this paper we employ
this approach to study the geometric phase and Hanny angle for the
generalized harmonic oscillator.  An apparent reason for this choice is
that this system admits the squeezed state as an exact solution, and
therefore is an ideal system for the study of time-dependent evolution of
a squeezed state. First, we will introduce the squeezed state approach or
squeezed state dynamics of the system. In this framework, the
time-evolution of a squeezed state can be described by an effective
Hamiltonian defined on an extended phase space and a differential equation
describing the phase change. The periodic and quasiperiodic solution of
the effective Hamiltonian correspond to the cyclic and quasicyclic
evolution of a squeezed state, respectively. This enables us to study the
cyclic evolution as well as the quasicyclic evolution of a squeezed state,
and their phase changes during the process.  In particular, we focus on a
class of cyclic states which are eigenstates of the Floquet operator and
of special interest in quantum physics. The cyclic states can be expressed
as a superposition of an infinite number of the squeezed states. Then,
their geometric phases are obtained analytically. It is demonstrated that
the quantal nonadabatic geometric phase is connected with the nonadiabatic
Hannay angle by a factor of $n+1/2$. In light of our discussions, the
quantal phase is found to be an area in the extended phase space swept out
by a periodic orbit, in both adiabatic and nonadiabatic cases. This gives
us a simple picture of  the geometric meaning of the quantal phase. 
   
The paper is organized as follows. In Sec. II, the squeezed state dynamics
is introduced for a generalized harmonic oscillator. An effective
Hamiltonian function and a differential equation describing the phase
change during the evolution will be given. Then, in subsequent three
sections, we are restricted to a specific choice of the periodic
parameters. We shall derive the non-adiabatic Hannay angle analytically
through the Lie transformation method in Sec. III. Sec. IV is devoted to
the discussions on the cyclic or quasicyclic motion of a squeezed state
and the phase change during the process. In Sec. V, we study the general
cyclic states - the eigenstates of the Floquet operator, and show the
connection between the geometric phase and the Hannay angle. In Sec VI,
our discussions  are generalized to a rather generic
situation without referring to the specific form of the parameters. The
paper concludes in Sec VII by discussions. 

\section{Squeezed  State Dynamics of a Generalized Harmonic Oscillator}

The squeezed state approach\cite{ZFG90,TF91} starts from the 
time-dependent variational principle (TDVP) formulation,
\begin{equation}
\delta \int dt\langle\Phi,t|i\hbar\frac{\partial}{\partial t} -
\hat H |\Phi,t\rangle = 0.
\end{equation}
Variation w.r.t $\langle\Phi,t|$ and $|\Phi,t\rangle$ gives rise to the
Schr\"odinger equation and its complex conjugate, respectively. 
In the squeezed state  
approach, the squeezed state is chosen as  the trial wave function. The 
squeezed state  is defined by the ordinary harmonic oscillator 
displacement operator acting on a squeezed vacuum state $|0\rangle$:
\begin{eqnarray}
|\Psi\rangle = exp(\alpha \hat a^{+} - \alpha^{*} \hat a)|\phi\rangle,
\nonumber\\
\quad |\phi\rangle = exp\left(\frac 1 2 (\beta{\hat a^{+2}} - \beta 
{\hat a}^2)\right)|0\rangle.
\end{eqnarray}
${\hat a}^{+}$ and ${\hat a}$ are boson creation and annihilation 
operator which satisfy the canonical commutation relation: $[\hat a,{\hat a}
^+] =1$. Squeezed states are of great
interest, offering many posssibe applications in 
diverse fields of quantum physics. Unlike  coherent
states, which have equal fluctuations in two directions of
the phase space and minimize the uncertainty product of Heisenberg's
uncertainty relation, the squeezed states
have less fluctuation in one direction
at the expense of increasing fluctuation in the other.

We define the coordinate and the momentum operators as:
\begin{eqnarray}
\hat p = i\sqrt{\frac{\hbar}{2}}({\hat a}^+ - \hat a),\nonumber\\ 
\hat q = \sqrt{\frac{\hbar}{2}}({\hat a}^+ + \hat a).
\end{eqnarray}
Thus we have
\begin{eqnarray}
 p \equiv \langle\Psi,t|\hat p|\Psi,t\rangle,\nonumber\\
    q \equiv \langle\Psi,t|\hat q|\Psi,t\rangle,\nonumber\\
   \Delta p^2 \equiv \langle\Psi,t|(\hat p-p)^2|\Psi,t\rangle=\hbar(\frac{1}{4G}+4\Pi^2G),\\
    \Delta q^2 \equiv \langle\Psi,t|(\hat q-q)^2|\Psi,t\rangle=\hbar G,\nonumber\\
   \langle\Psi,t|\hat q\hat p + \hat p\hat q|\Psi,t\rangle
   = 2qp + 4\hbar G\Pi.\nonumber
\end{eqnarray}
The canonical coordinate $(G,\Pi)$ was introduced by Jakiw and Kerman for
the quantum fluctuation, and its relations with $\beta$ in Eq.(2) is given
in Ref.\cite{JK79}. In fact, the squeezed state
$|\Psi,t\rangle$ is equivalent to the
following Gaussian-type state,
\begin{equation}
|\Psi,t\rangle=\frac{1}{(2G)^{1/4}}exp\left(\frac{i}{\hbar}(p\hat 
q-q\hat p)\right)
exp\left(\frac{1}{2\hbar}(1-\frac{1}{2G}+2i\Pi)\hat q^2\right)|0\rangle.
\end{equation}
From the TDVP, we can obtain the dynamical equations for the expectation 
values and the quantum fluctuations,
\begin{eqnarray}
    \dot q =\frac{\partial H_{eff}}{\partial p},\qquad
  \dot p = -\frac{\partial H_{eff}}{\partial q},\nonumber\\
    \hbar \dot G = \frac{\partial H_{eff}}{\partial \Pi},\qquad
    \hbar \dot \Pi = -\frac{\partial H_{eff}}{\partial G},
    \end{eqnarray}
where the dot denotes the time derivative.  The  effective 
Hamiltonian  $H_{eff}$ is defined on the extended space $(q,p,G,\Pi)$,
     \begin{equation}
     H_{eff} = \langle\Psi,t|\hat H|\Psi,t\rangle.
     \end{equation}
These equations give us a simple and clear picture about the motion of the
expectation values as well as the evolution of the quantum fluctuations. 

The time-dependent variational principle leaves an ambiguity of a      
time-dependent phase $\lambda (t)$. When the trial state $|\Psi,t\rangle$ is
transformed to $|\bar \Psi,t\rangle = e^{i\lambda(t)}|\Psi,t\rangle$, the
derived variational equation of motion remains invariant.
Therefore, we should fix the phase $\lambda(t)$ with the help of the 
Schr\"odinger equation,
\begin{equation}
\dot \lambda(t) = \langle\Psi,t|i\frac{\partial}{\partial t}|
\Psi,t\rangle
      -\frac{1}{\hbar}\langle\Psi,t|\hat H|\Psi,t\rangle.
\end{equation}
This phase is well defined for general {\it nonadiabatic} and 
{\it noncyclic} evolution of a squeezed state.
It represents a phase change  during
a time-evolution of squeezed state.
Obviously, the phase consists of two parts. The meaning of the second part is 
clear, it measures the time evolution. It is nothing but  
the {\it dynamical} phase  and can be rewritten as,          
\begin{equation}
\lambda_D(t) = -\frac{1}{\hbar}\int_0^t H_{eff} dt.
\end{equation}
The first part can be viewed as a difference between the {\it total} phase 
and the {\it dynamical} phase. We call it  {\it geometric} phase since it
just is the Aharanov-Anandan's phase
for  the case of cyclic evolution. From the expression
of the squeezed state, the geometric phase is equal to 
\begin{equation}
\lambda_G(t) =\int_0^t \left(\frac{1}{2\hbar}(p\dot q - q\dot p) -\dot\Pi 
G\right) dt. 
\end{equation}
It is easy to see from the above formula that  the evolution of
expectation value $(q,p)$ 
as well as the evolution of the quantum fluctuation $(G,\Pi)$ contribute
to the geometric phase. The contribution from the former one is
$\hbar$ dependent, while the contribution from quantum fluctuation
is $\hbar$ independent. 
For the case of cyclic squeezed state, the quantal phase is equal to a 
sum of the projective areas on the 
coordinates plane $(q,p)$  and fluctuation plane $(G,\Pi)$
swept out by a periodic orbit of the effective Hamiltonian.

In general situation, the squeezed state approach only gives an
approximate solution to the Schr\"odinger equation since the trial wave
functions are confined in a subspace of the Hilbert space. However, for a
Hamiltonian containing only quadratic terms, such as the generalized
harmonic oscillator that we shall discuss in the follows, it is easy to
prove that the method is accurate since the system admits squeezed state
as an exact solution. 

The Hamiltonian of the generalized harmonic oscillator 
takes the form,
\begin{equation}       
\hat H(q,p,t) =        
\frac 1 2 \left(a(t)\hat q^2 + b(t)\hat p^2       
+ c(t)(\hat q \hat p + \hat p \hat q)\right),       
\end{equation}
where $(\hat q,\hat p)$ are the quantum operators corresponding to       
the $(q,p)$, the conjugate variables of the phase space, and the       
real parameter ${a(t), b(t), c(t)}$ are periodic functions of the       
time with  common period $T$.
        
Applying the squeezed state  to this system, from Eq.(4) one can readily        
obtain an effective Hamiltonian in the extended phase space        
$(q,p;G,\Pi)$,
\begin{equation}        
H_{eff}(q,p;G,\Pi;t) = H_{cl}(q,p,t) + \hbar H_{fl}(G,\Pi,t),        
\end{equation}
where  
\begin{equation}
H_{cl} =         
\frac 1 2 \left(a(t) q^2 + b(t) p^2
+ 2 c(t) q p\right),
\end{equation}
describes the motion of the expectation values, and
\begin{equation}
H_{fl} = \frac 1 2 \left(a(t)G + b(t)(\frac{1}{4G} +4\Pi^2G) +
4c(t)G\Pi\right).
\end{equation}
depicts the evolution of the quantum fluctuations.

\section{Non-adiabatic Hannay Angle}

For the sake of simplicity, we first consider a specific choice of the
periodic parameters, namely, $a(t)=1+\epsilon \cos(\omega t),
b(t)=1-\epsilon \cos(\omega t) , c(t)=\epsilon\sin(\omega t)$. Our
discussions are restricted to the elliptic case, namely, $a(t)b(t) >
c^2(t)$, i.e. $\epsilon < 1$. For one period evolution the parameters
experience a circuit like $a+b=2; (a-1)^2+c^2=\epsilon^2$ in the parameter
space. We would like to point out that the results given below is by no
means limited to the specific chosen parameter form. As we shall show in
Sec.  VI that,  the main discussions can be  extended to the generic
case without referring to the form of parameters ${a(t),b(t),c(t)}$

The classical version of the generalized harmonic oscillator is described
by a classical Hamiltonian of the form $H_{cl}$. Without the perturbation
($\epsilon =0$), the phase plane of the system is full of the periodic
orbits with period $T_0=2\pi$. The $(q,p)=(0,0)$ is the unique fixed
point. When the perturbation is added ($1 > \epsilon > 0$), the $q=p=0$
remains a fixed point, however, all periodic orbits turn out to be
quasi-periodic.  In the Poincare section of $t=2n\pi/\omega$ ($n$ is an
integer), one observes a family of invariant tori. These tori are driven
by the Hamiltonian flow in such a way as to return to the original ones
after the time $T=2\pi/\omega$. This return leads to a classical angle
change, which is naturally separated into two parts: a dynamical part and a 
geometric part. The later one is connected with a phase-space area swept 
out over time $T$ and is named the non-adiabatic classical Hannay 
angle\cite{BH88}. 

The classical Hamiltonian can be rewritten in terms 
of the action-angle variables, namely,
$q=\sqrt{2I}\sin\phi,\quad p=\sqrt{2I}\cos\phi$,
\begin{equation}
H_{cl} = H_0 (I) + \epsilon H_1(I,\phi).
\end{equation}
where $H_0 = I, H_1 = -I\cos(\omega t + 2\phi)$. It is convenient to
employ the Lie transformation\cite{LL83} method to make a canonical
transformation, so that the new Hamiltonian $\bar H(\bar I)$ contains the
action variable only. Both the new Hamiltonian $\bar H$ and the generating
function $w$ are expanded in the power series as $\bar H
=\sum_{n=0}^{\infty}\epsilon^n \bar H_n$ and $w
=\sum_{n=0}^{\infty}\epsilon^n w_n$. For simplicity we shall introduce
some symbols first. Let $[\,,\,]$ represents a Poission Bracket; Lie
operator ${\cal L}_n$ is then defined by ${\cal L}_n = [w_n,\,]$;  and
operator ${\cal D}_0 = \frac{\partial}{\partial t} + [\,,H_0]$. Inserting
the series expansions and equating the terms of the same order of
$\epsilon$, one obtains a relation between the old and new Hamiltonian
functions (in what follows, our solutions are accurate to $\epsilon^2$),
\begin{eqnarray}
{\cal D}_0 w_1 = \bar H_1 -H_1,\nonumber\\
{\cal D}_0 w_2 = 2(\bar H_2 - H_2) - {\cal L}_1(\bar H_1 + H_1).
\end{eqnarray}
To the first order, we choose $\bar H_1$ to eliminate secularities in the
generating function $w_1$, and then solve for the generating function
$w_1$. To the second order, we substitute $w_1$ into the right hand side,
choose the $\bar H_2 $ to eliminate secularities in $w_2$, and solve for
$w_2$. Finally, we have the new Hamiltonian function in the form
\begin{equation}
\bar H(\bar I) = \bar I -\frac{\bar I}{\omega+2} \epsilon^2.
\end{equation}
The generating functions are $w_1=\frac{I\sin(\omega t + 2 \phi)}{\omega
+2}$ and $w_2 = 0$, respectively. The relation between the old variables
and the new variables is given by
\begin{equation}
(\phi, I)= {\cal T}^{-1} (\bar \phi, \bar I),
\end{equation}
where the transformation operator ${\cal T}^{-1} = 1 + \epsilon {\cal L}_1
+ \epsilon^2 ({\cal L}_2/2+{\cal L}_1^2/2)$. The above relations can be
expressed explicitly,

\begin{eqnarray} 
\phi = \bar\phi- \frac{\sin(\omega t + 2\bar
\phi)}{\omega +2}\epsilon + \frac{\sin(2\omega t +4\bar\phi)
}{2(\omega+2)^2} \epsilon^2 ,\nonumber\\ 
I = \bar I +\frac{2\bar I
\cos(\omega t+2\bar\phi)} {\omega+2}\epsilon +\frac{2\bar
I}{(\omega+2)^2}\epsilon^2. 
\end{eqnarray}
For this canonical transformation is explicitly time dependent, the new
Hamiltonian $\bar H$ differs from the old one $H_{cl}$ both in value and
in functional form.  Thus, we introduce a function $A$ to measure the
difference,
\begin{equation}
A(\bar \phi, \bar I ,t) = \bar H(\bar I) - H_{cl}\left(\phi(\bar\phi,\bar 
I,t), I (\bar \phi,\bar I,t),t\right) .
\end{equation}
Therefore  the classical non-adiabatic Hannay angle is
\begin{equation}
\Theta_H =\langle \int_0^T \frac{\partial A}{\partial \bar
I}dt \rangle_{\bar \phi_0},
\end{equation}
where the bracket denotes averaging around the invariant torus, $\langle
\cdots\rangle=\frac{1}{2\pi}\int_0^{2\pi}\cdots d\bar\phi_0$. 

With the help of Eqs. (15), (17) and (19), 
we finally arrive at the expression of the classical angle analytically,
\begin{equation}
\Theta_H = \frac{2\pi\epsilon^2}{(\omega +2)^2}.
\end{equation}
Obviously, this classical non-adiabatic Hannay angle is independent of
the action. 

\section{Cyclic and Quasicyclic  Squeezed  States}
        
At $\epsilon = 0$, as mentioned before, 
the system reduces to  the usual harmonic oscillator. The 
effective Hamiltonian can be expressed
in terms of the action-angle variables,
\begin{equation}        
H_{eff}(\epsilon=0) = I + 2\hbar (J + 1/2),        
\end{equation}
where,
\begin{equation}        
I = \frac {1}{2\pi}\oint pdq,\quad  J = \frac{1}{2\pi}\oint \Pi dG.        
\end{equation}
The transformations between $(q,p), (G,\Pi)$ and        
$(I,\phi), (J,\theta)$ take following  forms respectively,
\begin{eqnarray}
q=\sqrt{2I}\sin\phi,\nonumber\\
p=\sqrt{2I}\cos\phi,
\end{eqnarray}
and
\begin{eqnarray}        
G=2J+\frac 1 2 -\sqrt{2J(2J+1)}\cos\theta,\nonumber\\        
\Pi=\frac{\frac 1 2 \sqrt{2J(2J+1)}\sin\theta}        
{(2J+\frac 1 2 -\sqrt{2J(2J+1)}\cos\theta)}.        
\end{eqnarray}

One can clearly  see that, the motions of the two degrees of freedom
are degenerate. The phase space is full of the periodic orbits with
period $T_0=2\pi$ in $(q,p)$  and period $T_0/2$ in $(G,\Pi)$,
respectively. It is also  worthwhile  
pointing  out that, $q=p=0$, $G=1/2, \Pi=0$
is the unique fixed point of the system.

For the case of   
$\epsilon \ne 0$, the variables $(G,\Pi)$ are still decoupled from the
variables $(q,p)$.
In the fluctuation space $(G,\Pi)$,
all periodic orbits
degenerate to the quasi-periodic ones,
whereas the  fixed point
$G=1/2,\Pi=0$ bifurcates to a periodic orbit, whose expression 
can be easily  derived from Eq.(6) by using the method of   
power-series expansion,
$$ G_p(t)=\frac 1 2 -\frac{\cos(\omega t)}{\omega+2}\epsilon,$$
\begin{equation}
\Pi_p(t)=-\frac{\sin(\omega t)}{\omega+2}\epsilon.
\end{equation}
Then $(q=0,p=0;G_p(t),\Pi_p(t))$ is the unique solution with period T for
the effective Hamiltonian. It corresponds to a cyclic
squeezed state, whose
expectation values  keep fixed at the zero point, while its 
fluctuations ( the width of the wave packet) change periodically.
For this cyclic solution, its phase change  during one period can be 
evaluated by Eqs.(8-10),

\begin{equation}
\lambda_G = -\int_0^T \dot\Pi_p G_p dt=
 \int_0^T\Pi_p\dot G_p dt= -\frac{\pi\epsilon^2}
{(\omega+2)^2}. 
\end{equation}
\begin{equation}
\lambda_D =-\int_0^T dt 
H_{fl}\left(G_p(t),\Pi_p(t)\right)=-\frac{1}{2}\left(1+
(-\frac{2}{(\omega+2)}+\frac{2}{(\omega+2)^2})\epsilon^2\right) T.
\end{equation}

For the expectation values of the squeezed state keep fixed during the
cyclic evolution, the geometric phase is independent of the Planck
constant $\hbar$. Its value is equal to one half of the non-adiabatic
Hannay angle (Eq.(22)) except for a negative sign. 

Eq. (28) gives a unified expression for the Berry phase (in the
adiabatic limit $\omega \rightarrow 0$) and the Aharonov and Anandan's
phase (in nonadiabatic case). The geometric property of the former one
can be readily understood from its independent of the time history.
Whereas the geometric explanation of the nonadiabatic geometric phase has
to resort to the projective Hilbert space. Our expression Eq. (28) of the
geometric phase indicates a unified explanation, that is, the geometric
property of the phase rests with its being an area surrounded by a
periodic orbit in the fluctuation plane. 

Now we turn our attention to a class of quasiperiodic motion in the
extended phase space, that is, the fluctuations 
 $(G,\Pi)$ 
of a squeezed state   
keep staying on the  periodic orbit $(G_p(t),\Pi_p(t))$,
while its expectation value  moves along a quasiperiodic orbit. 
With the help of the Hamiltonian (17), this quasiperiodic solution is
described by
$$\bar I_q(t) =\bar I_0,$$  
\begin{equation}
\bar\phi_q(t) =\bar\phi_0 + (1-\frac{\epsilon^2}{\omega+2})t.
\end{equation}
(In all discussions, the initial time is set to be zero for convenience.)

Thus the  geometric phase change is obtained from Eq.(10)
\begin{equation}
\lambda_G(t)=\frac{1}{\hbar}\int_0^t 
I(t)
\dot \phi(t)
 dt + \int_0^t \Pi_p(t) \dot G_p(t) dt.
\end{equation}
where (from Eqs.(19) and (30))
$$I(t)\dot\phi(t)=I\left(\bar I_q(t),\bar \phi_q(t),t\right)
\dot \phi\left(\bar\phi_q(t),\bar I_q(t),t\right)$$
$$=\bar I_0 
-\frac{\bar I_0\omega}{\omega+2}
\cos(\omega t +2 t +2\bar\phi_0)\epsilon$$
\begin{equation}
+
\left(-\frac{2\bar I_0}{\omega+2}
+\frac{\bar I_0\sin(2\omega t+4t+4\bar \phi_0)}{\omega+2}
-\frac{\bar I_0\cos(2\omega t+4t+4\bar \phi_0)}
{\omega+2}+\frac{2\bar I_0}{(\omega+2)^2} \right)\epsilon^2.
\end{equation}

The dynamical phase is (from Eq.(9))

\begin{equation}
\lambda_D = -\frac{1}{\hbar}\int_0^t H_{cl}dt -\int_0^t H_{fl} 
\left(G_p(t),\Pi_p(t),t\right)dt.
\end{equation}
where (from Eqs.(13), (19) and (30)),
\begin{equation}
H_{cl} = \bar I_0 
- \frac{\bar I_0\omega}{\omega+2}
\cos(\omega t+2t+2\bar\phi_0)\epsilon+
\left(\frac{2\bar I_0}{(\omega+2)^2}-\frac{2\bar 
I_0}{\omega+2}\right)\epsilon^2. 
\end{equation}

\section{The general cyclic states: eigenstates of the Floquet operator}

For a time-periodic  Hamiltonian system, the Floquet theory
applies. A unitary time evolution operator  referring to one period T, the 
so-called Floquet operator $\hat U(T)$ is worthy of consideration,
since its eigenstates are obviously the cyclic states, and of great interest
 in physics.
The evolution operator satisfies $\hat U(mT) = \hat U^m (T)$.
One may ask: What is the form of the eigenstates? How about the
geometrical and dynamical phase of these cyclic states ? Is there
any connection between these phases with the Hannay's angle ? These 
questions will be answered in this section.

To make things simple, let us start by considering 
the case $\omega =1/s$, $s$ is a 
natural number. Now, we  construct a state as a superposition 
of infinite number of squeezed states, i.e.
\begin{equation}
|S_1\rangle = c\int_0^{2\pi}
 e^{\frac{i}{\hbar} \bar I_0 \bar\phi_0}|
\bar I_0,\bar\phi_0;G_0,\Pi_0\rangle d\bar\phi_0.
\end{equation}
where 
$|\bar I_0,\bar\phi_0;G_0,\Pi_0\rangle$
 represents a squeezed state centered
at $
q(\bar I_0,\bar \phi_0,t=0)
,p(\bar I_0,\bar \phi_0,t=0)$ (see Eqs.(19) and (25))
 with fluctuations $G_0,\Pi_0$; 
The $G_0,\Pi_0$ are chosen on  the unique periodic orbit 
$(G_0=G_p(t=0), \Pi_0=\Pi_p(t=0))$ (see Eq.(27)); $c$ is a 
normalization constant.

Acting the Floquet operator on the state, we have from Eqs. (30), (31) 
and (33),
\begin{equation}
\hat U(T)|S_1\rangle =
c\int_0^{2\pi}e^{\frac{i}{\hbar}\bar I_0 \bar \phi_0}
e^{i(\lambda_G(T)+\lambda_D(T))}
|\bar I_0,\bar\phi_0+(1-\frac{\epsilon^2}{\omega+2})T;G_0,\Pi_0\rangle
d\bar \phi_0.
\end{equation}
 From Eqs.(31) and (33), we obtain the expression of the phases,
\begin{equation}
\lambda_G(T) =
 \frac{\bar I_0 T}{\hbar}
\left(1-\frac{2\epsilon^2}{\omega+2}+
\frac{2\epsilon^2}{(\omega+2)^2}\right)
-
\frac{\pi\epsilon^2}{(\omega+2)^2}.
\end{equation}
and
\begin{equation}
\lambda_D(T)=\lambda_D^R=
-\left(\frac{\bar I_0}{\hbar}+\frac 1 2\right)\left(1+(\frac{-2}{\omega+2}+
\frac{2}{(\omega+2)^2})\epsilon^2\right)T.
\end{equation}

The minimum period of the trigonometry functions in the expression (32)
and (34) is $\frac{2s\pi}{2s+1}$, integration period in the calculation of
the phases is $2s\pi$.  Therefore, the terms containing $\bar\phi_0$
vanish. 

The  geometric phase is divided into following two parts,
\begin{equation}
\lambda_G(T) = \frac{\bar I_0 
T}{\hbar}\left(1-\frac{\epsilon^2}{\omega+2}\right)+ \lambda_G^R.
\end{equation}
where 
\begin{equation}
 \lambda_G^R=-\left(\frac{\bar I_0}{\hbar}+\frac{1}{2}\right)
\frac{2\pi\epsilon^2}{(\omega+2)^2}.
\end{equation}
As we will see later, the first part will compensate for a phase change
caused by the displacement of the expectation value, which makes the 
integrand having the same form as the original one under a
variable transformation.  
 
Making variable transformation 
$\bar \phi_0'=\bar \phi_0+(1-\frac{\epsilon^2}{\omega+2})T$,  
we have
\begin{equation}
\hat U(T)|S_1\rangle=ce^{i(\lambda_D^R+\lambda_G^R)}
\int_{
(1-\frac{\epsilon^2}{\omega+2})T
}^{2\pi+
(1-\frac{\epsilon^2}{\omega+2})T}
 e^{\frac{i}{\hbar} \bar I_0 \bar\phi_0'}|
\bar I_0,\bar\phi_0',G_0;\Pi_0\rangle d\bar\phi_0'.
\end{equation}
The integral    in above formula can be divided into three parts,
\begin{equation}
\int_{
(1-\frac{\epsilon^2}{\omega+2})T
}^{2\pi+
(1-\frac{\epsilon^2}{\omega+2})T}
 \cdots =
\int_{0}^{2\pi}\cdots
+
\int_{2\pi
}^{2\pi+
(1-\frac{\epsilon^2}{\omega+2})T}\cdots 
-
\int_{0
}^{
(1-\frac{\epsilon^2}{\omega+2})T} \cdots .
\end{equation}
The last two terms will cancel each other if and only if
$
 e^{\frac{i}{\hbar} \bar I_0 2\pi}=1$, which gives the 
condition for the state $|S_1\rangle$ being a cyclic state,
\begin{equation}
\bar I_0 = n\hbar .
\end{equation}
This is nothing but  the quantization rule without Maslov-Morse correction.

Under this condition, we get 
\begin{equation}
\hat U(T)|S_1\rangle = e^{i(\lambda_D^R +\lambda_G^R)}|S_1\rangle.
\end{equation}

Actually, the state $|S_1\rangle$ is an eigenstate of the Floquet
operator, $n$ is the state number. In comparison with Eq.(22), we finally
reach a simple relation between the geometrical phase and the non-adiabatic
Hannay angle,

\begin{equation}
\lambda_G^R=-(n+\frac 1 2)\Theta_H .
\end{equation}

Now we extend the above discussions to the general case, i.e. $\omega = r/s$,
where $r,s$ are co-primed natural numbers. First, let we consider the 
situation that $\hat U(mT)$ acts on the state $|S_1\rangle$,

\begin{equation}
\hat U(mT)|S_1\rangle = 
c\int_0^{2\pi}e^{\frac{i}{\hbar}\bar I_0 \bar \phi_0}
e^{i\lambda}
|\bar I_0,\bar\phi_0+(1-\frac{\epsilon^2}{\omega+2})mT;G_0,\Pi_0\rangle
d\bar \phi_0.
\end{equation}

Making  variable transformation   
$\bar \phi_0'=\bar \phi_0+(1-\frac{\epsilon^2}{\omega+2})mT$ and using
the  condition Eq.(43) one has,
\begin{equation}
\hat U(mT)|S_1\rangle = 
c\int_0^{2\pi}e^{\frac{i}{\hbar}\bar I_0 \bar \phi_0'}
e^{i\lambda'}
|\bar I_0,\bar\phi_0';G_0,\Pi_0\rangle
d\bar \phi_0'.
\end{equation}
where the phase $\lambda'$ can be naturally divided
into two parts $\lambda'=
\lambda_m^1 +\lambda_m^2(\bar \phi_0')$. 

The first term does not contain the variable $\bar\phi_0'$;
Those trigonometry functions in the phase that relates to the $\bar\phi_0'$ are 
included in the second term. Then we can derive the expression of the
first term through  a simple analysis,
\begin{equation}
\lambda_m^1 =
m(\lambda_G^R+
\lambda_D^R).
\end{equation}
Now we construct a new state,
\begin{equation}
|S_r\rangle = 
|S_1\rangle + ...+
e^{-i\lambda_m^1}\hat U(mT)|S_1\rangle
+ ...
e^{-i\lambda_{r-1}^1}\hat U((r-1)T)|S_1\rangle.
\end{equation}
Acting  the Floquet operator  on this state, we get,
\begin{equation}
\hat U(T)|S_r\rangle =
\hat U(T)|S_1\rangle + ...+
e^{-i\lambda_m^1}\hat U((m+1)T)|S_1\rangle
+ ...
+e^{-i\lambda_{r-1}^1}\hat U(rT)|S_1\rangle.
\end{equation}
Since  
\begin{equation}
\hat U(rT)|S_1\rangle = 
c\int_0^{2\pi}e^{\frac{i}{\hbar}\bar I_0 \bar \phi_0'}
e^{i(\lambda_r^1+\lambda_r^2(\bar \phi_0'))}
|\bar I_0,\bar\phi_0';G_0,\Pi_0\rangle
d\bar \phi_0'.
\end{equation}
From Eqs.(32) and (34),  we know that, the minimum  period of those 
trigonometry functions containing $\bar \phi_0$ is $\frac{2\pi s}{2s+r}$,
whereas the integral interval in calculating  phase is $rT=2\pi s$.
So the second term containing the $\bar \phi_0$ in the phase 
vanishes, and  
the last term in the Eq.(49) becomes,
\begin{equation}
e^{-i\lambda_{r-1}^1}\hat U(rT)|S_1\rangle=
e^{i(\lambda_r^1-\lambda_{r-1}^1)}|S_1\rangle.
\end{equation}
>From relation (48), one have $ \lambda_m^1-\lambda_{m-1}^1 =
\lambda_G^R+
\lambda_D^R$ for any $m$. 
Substituting this relation into Eq.(52) and then into Eq.(50) yields
\begin{equation}
\hat U(T)|S_r\rangle = e^{i(\lambda_D^R +\lambda_G^R)}|S_r\rangle.
\end{equation}
Then we conclude that the relation (45) still holds for the case $\omega = r/s$.
As $\omega$ is an irrational number, we can use a series of rational numbers to 
approach it. So we are convinced to  argue that, for any $\omega$ the
 geometric phase of the eigenvector of the Floquet operator is equal to 
$-(n+1/2)$ times the Hannay angle. 

The expression of the geometric phase $\lambda_G^R$ (see Eq.(40))
consists of two parts.
As discussed in Sec.IV, the
geometric meaning of the second term rests with its
representing an area in the fluctuation plane $(G,\Pi)$
swept out by a periodic
orbit. What is the meaning of the first term ? 
Now we should return to the coordinate plane $(q,p)$ and consider this problem
in the action-angle variables $(I,\phi)$. Let us consider following
differential 2-form which is preserved  under the canonical transformation
$(I,\phi) \rightarrow (\bar I, \bar\phi)$,
\begin{equation}
dI\wedge d\phi -d H \wedge dt = d \bar I \wedge \bar\phi -
d\bar H \wedge dt.
\end{equation}
Then the above formula can be rewritten as
\begin{equation}
dI\wedge d\phi  = d \bar I \wedge d\bar\phi -
d(\bar H-H) \wedge dt.
\end{equation}
Making an integration of the above equation for one period ($T$), and 
comparing it with  Eqs.(31) and (39), one will find 
immediately that the geometric
phase corresponds to the second term on  the right hand side of the 
above equation. 
The areas in the phase plane 
 $(I,\phi)$  and 
$(\bar I,\bar\phi)$ can be expressed respectively as, 
\begin{equation}
a_1 = \int_0^T I\dot\phi dt, \,\,\, a_2=\int_0^T \bar I \dot{\bar\phi} dt.
\end{equation}
Keeping in mind that the area meaning of the differential 2-form, one will find
that the first term of the geometric phase $\lambda_G^R$ represents the 
difference of the area through a canonical transformation. That is $
\langle a_1-a_2\rangle_{\phi_0}/\hbar$. Of course, an  averaging over the angle variable
 should be made.

\section{Extension to General Situation}

In the preceding sections, our discussions are restricted to a specific
choice of the periodic parameters $a(t), b(t)$ and $c(t)$. Explicit
perturbative results are obtained for the geometric phase and the Hannay
angle. In this section, we shall demonstrate that the main results
obtained in previous sections are also true for a generic situation. The
similar discussions can be done regardless of any concrete form of the
periodic parameters. 

In fact, whatever the form $a(t),b(t)$ and $c(t)$ have, under the elliptic
condition $a(t)b(t)>c^2(t)$, the phase plane of the expectation values
$(q,p)$ and the fluctuations $(G,\Pi)$ have the same topological structure
as that discussed in Secs. III and IV. One can clearly see that, in the
fluctuation plane $(G,\Pi)$, all motions are restricted on the invariant
tori except for a unique T-periodic solution $(G_p(t),\Pi_p(t))$. Whereas,
for the Hamiltonian system $H_{cl}$, the $(q=0,p=0)$ is obviously a fixed
point, other motions are quasi-periodic trajectories confined on the
invariant tori. This similarity in the topology of the phase structure
provides a basis for our generalization as given in the follows. 

Through a canonical transformation, $q=q(\bar I,\bar\phi,t),\quad p=p(\bar
I,\bar\phi,t)$, or inversely, $\bar I=\bar I(q,p,t), \bar\phi =
\bar\phi(q,p,t)$, the Hamiltonian $H_{cl}(q,p,t)$ can be transformed to a
new Hamiltonian $\bar H(\bar I,t)$ which does not contain the angle
variable $\bar \phi$. We still choose a state $|S_1\rangle$ in the same
form as in Eq.(35), and consider the situation that $\hat U(mT)$ acts on
the state $|S_1\rangle$,
\begin{equation}
\hat U(mT)|S_1\rangle = 
c\int_0^{2\pi}e^{\frac{i}{\hbar}\bar I_0 \bar \phi_0}
e^{i\lambda}
|\bar I_0,\bar\phi_0+
\int_0^{mT}\frac{\partial \bar H(\bar I_0,t)}{\partial \bar 
I_0}dt;G_0,\Pi_0\rangle d\bar \phi_0.
\end{equation}
where
\begin{equation}
\lambda = \lambda_D(mT) + \lambda_G(mT)
\end{equation}
\begin{equation}
 \lambda_D(mT)=-\frac{1}{\hbar}\int_0^{mT} H_{eff}dt
 \end{equation}
 \begin{equation}
 \lambda_G(mT) = \frac{1}{\hbar}\int_0^{mT}\frac {1}{2}
 (p\dot q -q\dot p)dt
 -\int_0^{mT}\dot\Pi_p G_pdt
 \end{equation}
$$\lambda_D(mT)=
\langle\lambda_D(mT)\rangle_{\bar \phi_0} + \{\lambda_D(mT) \}(\bar\phi_0),$$
\begin{equation}
\lambda_G(mT)=\langle\lambda_G(mT)\rangle_{\bar \phi_0}
+ \{\lambda_G(mT) \}(\bar\phi_0).
\end{equation}
where the symbols $\langle...\rangle_{\bar\phi_0}$
denotes the average over $\bar \phi_0$ as in Eq.(21);
$\{...\}(\bar\phi_0)$ represent
the terms relating to   $\bar\phi_0$.
\begin{equation}
\langle \lambda_G(mT) \rangle_{\bar\phi_0}
= \frac{m}{\hbar}\langle\int_0^T \frac 1 2 \left(p\dot q
 -q\dot p\right)dt\rangle_{\bar\phi_0}
 -m\oint G_p d\Pi_p .
 \end{equation}
Making variables transformation $\bar\phi_0' = \bar\phi_0 +
\int_0^{mT}\frac{\partial \bar H}{\partial \bar I_0}dt$ and under the condition
$\bar I_0 = n\hbar$, we have
\begin{equation}
\hat U(mT)|S_1\rangle = 
c e^{i\lambda_m^1}\int_0^{2\pi} e^{\frac{i}{\hbar}\bar I_0\bar\phi_0'}
e^{i\{\lambda_D(mT)\}(\bar\phi_0')+\{\lambda_G(mT)\}(\bar\phi_0')}|
\bar I_0,\bar \phi_0;G_0,\Pi_0\rangle d\bar\phi_0'
\end{equation}
where
\begin{equation}
\lambda_m^1 = m (\lambda_G^R+\lambda_D^R).
\end{equation}
and the geometric and dynamical parts take the forms as follows,
\begin{equation}
\lambda_G^R 
= \frac{1}{\hbar}
\left(\langle\int_0^T \frac 1 2 \left(p\dot q
-q\dot p\right)dt\rangle_{\bar\phi_0}
-\bar I_0\int_0^T\frac{\partial \bar H}{\partial \bar 
I_0} dt\right)
-\oint G_p d\Pi_p.
\end{equation}
\begin{equation}
\lambda_D^R = -\frac{1}{\hbar}<\int_0^T H_{eff}dt>_{\bar\phi_0}
\end{equation} 
The motion of the expectation values $(q,p)$
confined on the invariant torus $\bar I_0$ is quasi-periodic. Ergodicity
of the motion guarantees that the temporal average is equivalent to
the spatial average supposing that the time is long enough.
 Then, in light of the ergodicity principle
  we can choose an integer $r$, which is large
enough so that the terms containing $\bar\phi_0$ in phase expression
$\lambda$ in Eq.(57) almost vanish. Then, similar to Eq. (49), we
construct a state $|S_r\rangle$ and readily prove that the relation (53)
still holds. Now let us see the meaning of the geometric phase
$\lambda_G^R$ expressed by (65). Similar to (54) and (55), we consider the
following differential 2-form which is preserved under the canonical
transformation, i.e. 
\begin{equation}
dp\wedge dq -   d\bar I\wedge d\bar \phi = -d(\bar H -H_{cl})\wedge dt .
\end{equation}
Similar to  discussions in the last paragraph of the former section,  
 let us first
 make an integration of the above equation for one period
(T),
 then
average over the variable $\bar\phi_0$.
Keeping in mind that the area meaning of the differential 2-form,
one will find immediately that the term bracketed in the expression
of the geometric phase (65) corresponds to the left hand side of
the above equation, whereas the right hand side will equal to
$n\hbar$ times the Hannay's angle (refer to (21)).
1/2 relation between the first  term in (65) and the classical angle is   
given by Ge and Child\cite{GC97} and verified by our explicit    
perturbative results in former sections.
Thus,
we arrive at the  same  conclusion as is shown in Eq.(45).
   
\section{Conclusions and Discussions}
 
In this paper, the nonadiabatic geometrical phase and the Hannay angle of
the cyclic evolutions for a generalized harmonic oscillator are studied by
using the squeezed state approach. It is demonstrated that the squeezed
state approach is a very powerful method, which enables us to obtain
analytically and explicitly the geometrocal phase and the Hannay angle. 

In the framework of the squeezed state, the time-dependent evolution of a
squeezed state can be described by an effective Hamiltonian on an extended
phase space and a differential equation describing the phase change. The
periodic and quasiperiodic solution of the effective Hamiltonian system
corresponds to the cyclic and quasicyclic evolution of a squeezed state,
respectively. Then, we found a unique cyclic squeezed state, whose
expectation value keeps staying at the zero point, while the fluctuations
change periodically. We obtain a unified expression of its geometrical
phase for the adiabatic as well as nonadiabatic case. The geometric
property lies in its equality to an area on the extended phase space swept
out by a periodic orbit. 

A special class of cyclic states of the system are of great interest in
physics.  They are the so-called the Floquet states -eigenvectors of the
Floquet operator. These cyclic states can be expressed as a superposition
of an infinite number of the squeezed states. Their geometric phases are
obtained analytically. The quantum phase can be interpreted as a sum of
the area difference on the expectation value plane through the cannonical
transformation and the area on the quantum fluctuation plane swept out by
a periodic orbit. This explanation provides a unified picture of the
geometric meaning of the quantal phase for the adiabatic case as well as
the nonadiabatic case. 

We have also discussed the classical version of the system. The
nonadiabatic Hannay angle is obtained by employing Lie transformation
method, which is found to be independent of the action. The geometrical
phases of those cyclic states are equal to $-(n+1/2)$ times the Hannay
angle.  In the adiabatic limit, our $n+1/2$ relation is identical to the
elegant formula of Berry\cite{HB85}. However, the semiclassical
approximation has not been envoked. Therefore, we believe that, in
addition to the semiclassical method, the squeezed state approach provides
an alternative way to bridge the classical and quantum world. 

An interesting example is given by $n=0$, i.e. the ground state of the
Floquet states. It corresponds to the unique cyclic squeezed state
mentioned above. The geometrical phase of this cyclic state resulting only
from the periodic evolution of the fluctuations' part, is equal to one
half of the classical angle. This is just what obtained by Ge and Child
\cite{GC97}.

\section*{Acknowledgments}

We would like to thank Profs. Shi-Gang Chen, Lei-Han Tang, and Wei-Mou
Zheng for helpful discussions and comments. This work was supported in
part by the grants from the Hong Kong Research Grants Council (RGC) and
the Hong Kong Baptist University Faculty Research Grants (FRG).



\begin{thebibliography}{99}
\bibitem{Berry84}
M. V. Berry, Proc. Roy. Soc. London, A392, 45-57 (1984)
and references in 
 {\it Geometric Phase in Physics}, ed. A. Shapere and F. Wilczek.
(1989), World Scientific.
\bibitem{HB85} J. H. Hannay, J. Phys. A {\bf 18}, 221 (1985);
M. V. Berry, J. Phys. A {\bf 18}, 15 (1985).
\bibitem{AA87} 
Y. Aharonov and J. Anandan, Phys. Rev. Lett. {\bf 58}, 1593, (1987).
\bibitem{BH88} 

M. V. Berry and J. H. Hannay, J. Phys. A {\bf 21}, L325 (1988).
\bibitem{GC97} 
Y. C. Ge and M. S. Child, Phys. Rev. Lett. {\bf 78}, 2507 (1997).
\bibitem{PS94}
A. K. Pattanayak and W. C. Schieve, Phys.Rev.Lett. {\bf 72}, 2855 (1994) 
\bibitem{ZF95}
W. M. Zhang and D. H. Feng, Phys. Rep. {\bf 252}, 1 (1995), and references 
therein.
\bibitem{LS97}
W. V. Liu and W. C. Schieve, Phys. Rev. Lett. {\bf 78}, 3278 (1997)
\bibitem{HLLZ98}
B. Hu, B. Li, J. Liu and J. L. Zhou, Phys. Rev. E {\bf 58}, 1998 (in press)
\bibitem{ZFG90}
W. M. Zhang, D. H. Feng and R. Gilmore, Rev. Mod. Phys. {\bf 62}, 867 (1990).
\bibitem{TF91} 
Y. Tsui and Y. Fujiwara, Prog. Theor. Phys. {\bf 86}, 443 (1991).
\bibitem{JK79}
R. Jackiw and A. Kerman, Phys. Lett. A {\bf 71}, 158 (1979).
\bibitem{LL83}
A. J. Lichtenberg and M. A. Lieberman, {\it Regular and Stochastic Motion},
P123, Springer-Verlag, (1983)
\end{thebibliography}
\end{document}